# Justification of Power-Law Canonical Distributions Based on Generalized Central Limit Theorem

Sumiyoshi Abe[1] and A. K. Rajagopal[2]

[1]*College of Science and Technology, Nihon University,*

*Funabashi, Chiba 274-8501, Japan*

[2]*Naval Research Laboratory, Washington, DC 20375-5320*

A self-consistent thermodynamic framework is presented for power-law canonical distributions based on the generalized central limit theorem by extending the discussion given by Khinchin for deriving Gibbsian canonical ensemble theory. The thermodynamic Legendre transform structure is invoked in establishing its connection to nonextensive statistical mechanics.

Systems exhibiting power-law behavior in their distributions are often encountered in nature. Until very recently, understanding of such properties based on the principles of Gibbsian canonical ensemble theory has not been satisfactory due to the exponential form of its distributions. Gibbsian canonical ensemble theory has been variously elaborated by several approaches such as the method of steepest descents, counting algorithm, maximum entropy principle, and the ordinary central limit theorem. (See Refs. [1,2] for a comprehensive account.) Among all these approaches, the one based on the central limit theorem seems to be the most satisfactory with the least number of underlying assumptions [3]. In this Letter, we present the derivation of the non-Gibbsian power-law canonical distributions based on the generalized central limit theorem of Lévy and Gnedenko [4]. In the spirit of Khinchin [3], we thus establish the clear role played by probability theory in statistical mechanics of physical systems obeying power-law distributions.

Suppose that a given system is composed of a large number of subsystems with energies $\{\varepsilon_i = E_i/B_N > 0\}_{i=1,2,\ldots,N}$, where $B_N$ is a positive $N$-dependent factor which will be determined subsequently. The total energy is assumed to be the sum $E = \varepsilon_1 + \varepsilon_2 + \ldots + \varepsilon_N$. Here, it is useful to recall the ordinary central limit theorem. If each of $E_i$'s obeys a common distribution $f(E_i)$ with ordinary finite second moment $\langle E_i^2 \rangle_1 = \int_0^\infty dE_i\, E_i^2\, f(E_i)$, then its $N$-fold convolution, $B_N(f*f*\ldots *f)(B_N E)$, approaches the Gaussian distribution in the limit of large $N$, where $(f*g)(x) \equiv \int_0^x dx'\, f(x-x')\, g(x')$. This property was exploited by Khinchin to establish Gibbsian canonical ensemble theory within the framework of probability theory. However, there are distributions for which all of the ordinary moments, $\langle E_i^n \rangle_1 = \int_0^\infty dE_i\, E_i^n\, f(E_i)$ $(n=1,2,3,\ldots)$, are divergent. In spite of this property, there still exists the limiting distribution, to which the $N$-fold convolution of the distribution of this type may converge [4]. Mathematically, this statement is called the generalized central limit theorem and the associated limiting distribution is referred to as the Lévy-stable distribution, denoted here by $F_\gamma(E_i)$. The latter is the case which leads to the power-law distributions of concern in the present work. The characteristic function of the Lévy distribution in the half space is known to be of the form [4]:

$$\chi_\gamma(t) = \int_0^\infty dE_i \, e^{iE_i t} \, F_\gamma(E_i)$$

$$= \exp\left\{-a|t|^\gamma \exp\left[i\,\mathrm{sgn}(t)\frac{\theta\pi}{2}\right]\right\}, \tag{1}$$

where $a$ is a positive constant, $\gamma$ the Lévy index lying between $(0,1)$, $\theta$ a constant satisfying $|\theta| \leq \gamma$, and $\mathrm{sgn}(t) = t/|t|$ the sign function of $t$. This is an invariant distribution in the sense that if each of the $E_i$'s obeys $F_\gamma(E_i)$, then their $N$-fold convolution $B_N(F_\gamma * F_\gamma * \text{L} * F_\gamma)(B_N E)$ has the same characteristic function as that of the original $F_\gamma(E_i)$, i.e., $\chi_\gamma^{(N)}(t) \equiv [\chi_\gamma(t/B_N)]^N = \chi_\gamma(t)$, if $B_N$ is chosen to be

$$B_N = N^{1/\gamma}. \tag{2}$$

$F_\gamma(E_i)$ is found to exhibit the following power-law behavior for large values of $E_i$:

$$F_\gamma(E_i) \sim E_i^{-1-\gamma}. \tag{3}$$

Now, we consider a power-law distribution

$$f(E_i) = \frac{1}{z(\beta)}[\xi(\beta) + E_i]^{-s} \quad (1 < s < 2), \tag{4}$$

$$z(\beta) = \frac{\xi^{1-s}(\beta)}{s-1}, \tag{5}$$

where $\xi(\beta)$ is a positive constant depending on a certain parameter $\beta$. The range of $s$ is determined in such a way that $f(E_i)$ is normalizable and all of the ordinary positive integer moments are divergent. The logarithm of the characteristic function of the $N$-fold convolution of this distribution with the same choice of $B_N$ as in eq. (2) is calculated to be

$$\ln \chi^{(N)}(t) = -a|t|^{s-1} \exp\left[i\,\mathrm{sgn}(t)\frac{\theta\pi}{2}\right]$$

$$+(s-1)\,\xi^{s-1}|t|^{s-1} \int_0^\infty dy\,(\exp[i\,y\,\mathrm{sgn}(t)]-1)\left[\frac{1}{\left(y+\dfrac{\xi|t|}{B_N}\right)^s} - \frac{1}{y^s}\right]. \quad (6)$$

Here, $a$ and $\theta$ are respectively given by

$$a = (s-1)\,\xi^{s-1}(\beta)\,|L(s-1)| > 0, \quad (7)$$

$$\theta = 1-s, \quad (8)$$

where [4] (p. 169)

$$L(\alpha) = \int_0^\infty \frac{dy}{y^{1+\alpha}}(e^{-y}-1) < 0 \quad (0<\alpha<1). \quad (9)$$

The last term in eq. (6) is the correction term, which vanishes in the limit $N \to \infty$. In deriving the above result, we have compared eq. (6) with eq. (1) and have found the Lévy index to be

$$\gamma = s-1. \quad (10)$$

To completely characterize the limiting distribution $F_\gamma(E)$, it is necessary to relate the constant $a$ to a statistical physical quantity. Note that, in contrast to Khinchin's discussion based on the ordinary central limit theorem, here we have no finite ordinary moments of the distribution. This calls for a different approach to the determination of the constant $a$ appearing in eq. (6). Our strategy is, therefore, to fix the constant $\xi$ by some means using the distribution of a single subsystem.

To this end, we observe that although the ordinary first moment, $\langle E_i \rangle_1$, is divergent the generalized first moment, $\langle E_i \rangle_q \equiv \int_0^\infty dE_i \, E_i \, P_q(E_i)$, defined in terms of the escort distribution [5] associated with $f(E_i)$, that is,

$$P_q(E_i) \equiv \frac{f^q(E_i)}{\int_0^\infty dE_i \, f^q(E_i)}, \qquad (11)$$

is designed to be finite if $q$ is chosen to be

$$q = 1 + \frac{1}{s}. \qquad (12)$$

This choice is dictated by the minimal requirement from the finiteness of $\langle E_i \rangle_q$. Let us calculate the denominator in eq. (11) as well as $\langle E_i \rangle_q$:

$$c_q \equiv \int_0^\infty dE_i \, f^q(E_i) = \frac{(s-1)^q}{s} \xi^{1-q}(\beta), \qquad (13)$$

$$u_q \equiv \langle E_i \rangle_q = \frac{\xi(\beta)}{s-1}. \qquad (14)$$

Therefore, we clearly see how the constant $a$ in eq. (7) is expressed in terms of $u_q$, which is a physical quantity, namely the generalized internal energy. From eqs. (13) and (14), it also follows that

$$c_q = \frac{s-1}{s} u_q^{1-q}. \qquad (15)$$

In order to develop the connection with thermodynamics, we regard $\beta$ and $u_q$ as a conjugate pair in the thermodynamic Legendre transform structure [6]. To do this, we define an entropy functional, $S_q[f]$, which satisfies

$$\frac{\partial S_q}{\partial u_q} \equiv \beta, \qquad (16)$$

$$-\frac{\partial \Gamma_q}{\partial \beta} \equiv u_q, \qquad (17)$$

where

$$\Gamma_q \equiv S_q - \beta u_q. \qquad (18)$$

This requirement determines the relation between $\xi$ and $\beta$ appearing in eq. (4), once the form of the entropy $S_q[f]$ is specified. At this juncture, we examine the following form

$$S_q[f] = \frac{1}{1-q}\left(\sigma^{1-q} c_q - 1\right), \qquad (19)$$

where we have introduced a scale $\sigma$ to make $S_q[f]$ dimensionless. Then, using eq. (16), we obtain

$$\beta = \frac{s}{s-1} \frac{u_q^{-q}}{\sigma^{1-q}}. \qquad (20)$$

From eqs. (18) and (20), we verify that equation (17) indeed holds, as it should do.

We complete our discussion by pointing out that actually the entropy in eq. (19) and the associated distribution of a single subsystem in eq. (4) are respectively the Tsallis nonextensive entropy [7] and its optimal distribution under the constraint on the generalized first moment with respect to the escort distribution in eq. (11) [8]. We mention that similar studies on convergence properties in both the full and half spaces have recently been reported in the context of nonextensive statistical mechanics in Refs. [9,10].

In conclusion, we have constructed a self-consistent thermodynamic framework for power-law canonical distributions by making use of the generalized central limit theorem in a similar manner to that given by Khinchin for deriving Gibbsian canonical ensemble theory. In establishing a connection between generalized canonical ensembles with power-law distributions and nonextensive statistical mechanics, we have utilized the thermodynamic Legendre transform structure, which is an underlying principle common in both extensive and nonextensive statistical mechanics.

One of us (S.A.) was supported by the GAKUJUTSU-SHO Program of College of Science and Technology, Nihon University. He thanks the warm hospitality of the Naval Research Laboratory, Washington, DC, which made this collaboration possible. The other author (A. K. R.) acknowledges the partial support from the US Office of Naval Research.